\documentstyle[12pt]{article} 
\begin{document}
\title{Spiral Molecular Front in Galaxies:
Quick Transition from Atomic to Molecular Hydrogen in Spiral Arms}
\author{Makoto HIDAKA, \& Yoshiaki SOFUE\\
{\it Institute of Astronomy, University of Tokyo,
    Mitaka, Tokyo, 181-0015, Japan} \\
	E-mail: sofue@ioa.s.u-tokyo.ac.jp}

\def\fmol{f_{\rm mol}}
\def\vrot{v_{\rm rot}}
\def\kms{ km s$^{-1}$ }
\def\htwo{H$_2$}
\def\be{\begin{equation}}
\def\ee{\end{equation}}
\def\ref{\hangindent=1pc \noindent}

\maketitle 

\begin{abstract}
We derived a two-dimensional map of the molecular fraction, $\fmol$,
(ratio of the molecular gas density to that of total gas) in the spiral
galaxy M51, and examined the behavior of molecular fronts (MF), where
MF represents the place where $\fmol$ changes drastically from nearly 
zero to unity and vice versa.
We show that the MF phenomenon occurs not only radial, but also in the
azimuth direction through the spiral arms, and $\fmol$ changes rapidly
in the arm-to-inter-arm transition regions.
The existence of the azimuthal MF indicates that the atomic gas (HI) is quickly
transformed to molecular gas (H$_2$) during the passage through spiral arms.
We performed a numerical simulation of MF based
on an HI-to-\htwo\ phase transition theory, and reproduced the observations.
We estimated a azimuthal scale length of the transition to be less than 
200 pc, corresponding to a time scale of $\sim$ 2 Myr for \htwo\ gas formation.
The azimuthal width of a molecular arm is estimated to be at most
2.5 kpc, where the gas can remain in molecular phase for about 25 Myr.

Kew words: galaxies: spiral --- galaxies: ISM --- galaxies: cluster of --- 
Molecular gas --- HI gas

\end{abstract}

\section{Introduction}

The neutral (HI) and molecular hydrogen (H$_2$) gases are the two major 
components of the interstellar matter.
Elmegreen (1993) has proposed a simple model to predict
the molecular fraction by assuming a quasi-static  equilibrium between the
two phases in individual interstellar clouds.
However, the dynamical transition between HI and \htwo\ gas phases are not well
understood in real galaxies: Are the GMCs (giant molecular clouds) long-lived
clouds living for several galactic rotations crossing the spiral arms for
many times, coexisting with the HI gas?
Alternatively, are they formed quickly from HI gas when the inter-arm HI
encounters the arms, and does molecular gas return to HI again, when going 
out of the arms?

This question can be answered by studying the distribution of molecular
fraction, the ratio of molecular-gas density to the total gas density,
in the gas disk of spiral galaxies.
Global variations of the molecular fraction in disk galaxies have been
obtained by our earlier works (Sofue et al 1995; Honma et al 1995),
applying the Elemegreen's theory of atomic-to-molecular gas transition.
We showed that the inner gas disk is dominated by molecular gas, while the
outer disk by HI, and the transition between HI and \htwo\ occurs in a narrow
annulus in the disk, across  which the molecular fraction varies drastically.
We call this transition region the ``molecular front (MF)''.
Such front phenomenon was found also in the direction perpendicular to the
galactic plane, so that the \htwo\ disk is sandwiched by a fat HI disk
(Imamura and Sofue 1997).
However, these analyses do not tell us about dynamics and time scale of
the HI-to-\htwo\ transition.

In the present paper, we investigate the MF phenomenon
across spiral arms by obtaining a two-dimensional distribution of
molecular fraction in a galactic disk.
Azimuthal variation of molecular fraction across spiral arms will provide
us with information about dynamical evolution of the HI-\htwo\ phase
transition and molecular gas formation in spiral arms.

\section{Molecular Front}

\subsubsection{Radial and Vertical Molecular Front}
Many observations of the neutral hydrogen and the molecular gas have been
carried out, and have revealed two important aspects.
First, the neutral hydrogen gas is broadly distributed in the
disk, and is often extended outside the optical disks,
while it is deficient in the central regions.
On the other hand, molecular gases are concentrated in the
central a few kpc regions.
The radial variation of the distributions of HI and \htwo\ gases have been
quantified by analyzing the molecular fraction as a function of the
galactco-centric distance.
The molecular fraction, $\fmol$, is defined by
\begin{equation}
\fmol = \frac{\rho (\rm{H}_2)}{\rho (\rm{HI}) + \rho (\rm{H}_2)}
= \frac{2 \times n(\rm{H}_2)}{ n(\rm{HI}) + 2 \times n(\rm{H}_2)}.
\end{equation}
 Sofue et al. (1995) have investigated radial variation of the molecular
 fraction in several spiral galaxies.
 They found  that the molecular fraction is almost unity in the central
 a few kpc region, and decreases suddenly at a certain radius to almost zero,
 beyond which the gas is almost totally in the HI phase.
Such a step-like variation of the molecular fraction is called the
molecular front.

Honma et al. (1995) modeled the molecular front by applying the phase
transition theory of Elmegreen (1993).
According to this theory, the molecular fraction is determined
by three parameters: interstellar pressure $P$, UV radiation field $U$ ,
and metallicity $Z$.
Honma et al showed that, if these three parameters are approximated by
exponential functions of galacto-centric radius, the observed molecular
front in the radial direction can be well reproduced.
They applied this model to several real galaxies, and obtained good
coincidence with observed data.

Spiral galaxies have density waves and spiral arms.
As the spiral pattern speed is much slower than the gas's rotation,
the gases collide with the spiral potential, resulting in galactic shock
waves, where the gas density and pressure increase suddenly.
The compression results in star formation, leading to increase of
the radiation field intensity.
Thus, the MF parameters, $(P, U, Z)$,  will vary as a function of azimuthal
angle, which may results in another type MF spatially correlated with the
spiral arms.
In order to investigate the existence of such  spiral MF,
we first examine the HI and CO-line data from the literature,
and construct a 2D map of the molecular fraction in M51.
Then, we perform numerical simulation using the transition
model established by Elmegreen (1993), and compare with the observations.

\subsection{Spiral Molecular Front in M51: Azimuthal Variation of 
Molecular Fraction}

For obtaining two-dimensional map of molecular fraction, we make use of the
data from the literature that provide sufficient angular resolution in both
HI and CO. Here, we choose the `grand-design' spiral galaxy NGC 5194 (M51),
which was also studied in Honma et al. (1995).
We use the $^{12}$CO ($J=1-0$) imaging data obtained with the Nobeyama 45-m
telescope (Nakai et al 1994) with the beam width of $16''$ and grid 
spacing of $15''$, corresponding to an angular resolution of $24''$, 
and the HI data obtained with the VLA at $34''\times 34''$ resolution 
from Rots et al. (1990).
Integrated-intensity maps of HI and CO are shown in Fig. 1 at the same
resolution of $34''$.
These two maps reveal the typical behaviors of atomic and molecular hydrogen
gases: HI is extended broadly in the disk, having central deficient,
while CO is concentrated in the central region.

---- Fig. 1 ----

We calculate the column density of HI and \htwo\ from the integrated intensity,
adopting the following relation between the column density and intensity:
\begin{eqnarray}
n(\rm{HI}) &=& C(\rm{HI}) \times I(\rm{HI}) , \\
n(\rm{H}_{2}) &=& C(\rm{CO}) \times I(\rm{CO}),
\end{eqnarray}
where $C(\rm{HI})$ and $C(\rm{CO})$ are the conversion factors.
The value of $C(\rm{HI})$ can be easily calculated as
$1.82 \times 10^{18}$ cm${}^{-2}$ K${}^{-1}$ km${}^{-1}$ sec.
In order to estimate the amount of the total molecular gas mass from CO
intensity, we adopt a conversion factor of
$C(\rm{CO}) = 1.1 \times 10^{20}$ cm${}^{-2}$ K${}^{-1}$ km${}^{-1}$ sec
from Arimoto et al. (1996),  and assume no radial dependence of the conversion
factor.
The difference of the angular resolution was corrected, so that both HI and
CO maps have the same resolution of $34''$, and the grid separation was
corrected by using bicubic interpolation. Using Eq. 1, we calculated the
 molecular fraction, and show a two dimensional distribution map of
the molecular fraction in Fig. 2.

---- Fig. 2 ----

Fig. 2 shows that the molecular fraction is almost unity in the central
$ r < 60''$ region, or at $r < 2.8$ kpc.
Toward the northwest and southeast,
where the bisymmetric spiral structure is not seen clearly in both HI and CO
distribution, the molecular fraction decreases drastically to zero
at  $r > 110''$, or $r > 5.1$ kpc.
This sudden radial decrease of the molecular fraction confirms the
radial MF at  $r \sim 5$ kpc.

In the northeast and  southwest, where the molecular spiral
arms are prominent, a high-$\fmol$ region extends until $\sim 7$ kpc.
In the southeast direction, $\fmol$ suddenly decreases to 0.6 at
$r \sim 60^{\prime\prime}$, and increases again to a value greater than 0.9,
and finally decreases to zero.
This re-increase of the molecular fraction results from the spiral arms,
as already mentioned by Kuno et al. (1997).

Fig. 2 shows  that the molecular front exists along
the spiral arms in addition to the global radial variation.
This implies that the circularly flowing gas experiences quick transition
from atom to molecule by encountering the spiral arms,
and from molecule to atom when going out from the spiral arms.
However, the present angular resolution is not high enough to investigate the
scale length of the variation across the arms in more detail.
We can estimate only the upper limit of the scale length, which
is about 700 pc in the azimuthal direction.
Since the rotation velocity is about 200 km/sec (Kuno \& Nakai 1997)
and pattern speed of the density waves will be $\sim 20$ \kms\ kpc$^{-1}$,
the upper limit of time scale of atom-to-molecule
and molecule-to-atom phase transitions is estimated to be several Myr.

\section{Simulation of the Molecular Front}

\subsection{Numerical Methods}

In previous section, we have shown the existence of MF in
azimuthal direction. However, because of poor angular resolution, we could
not estimate the exact MF scale length in the arms.
Here, we try to estimate the scale length
by numerical simulations.
The azimuthal variances of the gas
density is calculated using the hydrodynamical code, not taking
into account the self-gravity of the gas. The variation of metallicity and
radiation field is assumed to have an axisymmetric exponential distribution,
as adopted by Honma et al. (1995), but no azimuthal variation is assumed.

For calculating the gas dynamical behavior of the disk, we use the freely
downloadable and usable hydrodynamical code, VH-1 (Blondin \& Lufkin 1993).
This is a multidimensional hydrodynamics code for an ideal compressible fluid
written in FORTRAN, developed by the numerical astrophysics group at the
University of Virginia on the bases of the Piecewise Parabolic Method
Lagrangian Remap (PPMLR) scheme of Colella \& Woodward (1984).
The PPMLR has the advantage of
maintaining contact discontinuities without the aid of a contact steepener,
and is good to be applied to a galactic scale hydrodynamical simulation.
The code does not include self-gravity, artificial viscosity,
variable gamma equation of state, and radiative heating and/or cooling.
Hence,  we assume that the interstellar gas is ideal, inviscid, and
compressible.

\subsubsection{Gravitational Potential}
In order to simulate the evolution of spiral structure in the gas disk,
we assume a given gravitational potential, which comprises the following
two terms:
(i) a static axisymmetric potential, and
(ii) a nonaxisymmetric, rotating bar potential.
The self-gravity of the gas is not taken into account.
The potential is expressed by
\begin{equation}
\Phi (R, \phi ) = \Phi_0 (R) + \Phi_1 (R, \phi ).
\end{equation}
We adopt a ``Toomre disk'' (Toomre 1981) potential for the axisymmetric
component as given by:
\begin{equation}
\Phi_0 (R) = - \frac{c^2}{a} \frac{1}{(R^2 + a^2)^{1/2}}, \label{eq:4.5}
\end{equation}
where $ a $ is the core radius, and $c = v_{\rm{ max}} (27/4){}^{1/4} a$
with $v_{\rm{ max}}$ being the maximum rotation velocity.
Through our numerical simulation, we fix the core radius
and maximum circular velocity to be $a = \sqrt{2}$ kpc and $v_{\rm {max}}
= 200$ \kms.

The nonaxisymmetric potential is taken from Sanders (1977), assuming rigid
rotation at a pattern speed $\Omega_p$, which has the form
\begin{equation}
\Phi_1 (R, \phi ) = \varepsilon \frac{aR^2}{(R^2 + a^2)^{3/2}} \Phi_0(R)
\cos 2\left(  \phi - \Omega_p t \right), \label{eq:4.6}
\end{equation}
where $\varepsilon$ is the strength of the bar of the order of
$\varepsilon = 0.15$.
Spiral shocked arms of gas are produced by this potential.

\subsubsection{Initial Condition}
Initially, we set $256 \times 256$ two-dimensional cells corresponding to
$12.8 \times 12.8$ kpc field, setting the field center at
the coordinates origin.
The initial number density of the gas is taken to be 5 cm${}^{-1}$ for the 
inner disk at $R \le 8$ kpc, and 1 cm${}^{-1}$ at $R> 8$ kpc.
Initial rotation velocity of each gas cell is set so as for the centrifugal
force to balance the gravitation.
The bar pattern speed $\Omega_p$ is taken to be 23 \kms kpc$^{-1}$, and
the strength of the bar $\varepsilon$ is taken to be 0.10.

\subsection{Elmegreen's Parameters}
We specify three parameters: interstellar pressure $P$, radiation field
strength $U$, and metallicity $Z$.
Following Honma et al. (1995), we represent $U$ and $Z$ by the
following equations.
\begin{eqnarray}
\frac{U}{U_\odot} &=& \exp \left( - \frac{R - R_{U_s}}{R_{U}} \right),\\
\frac{Z}{Z_\odot} &=& \exp \left( - \frac{R - R_{Z_s}}{R_{Z}} \right).
\end{eqnarray}
Here, $R_{U}$ and $R_{Z}$ are the scale radii of the radiation field
and metallicity, respectively, and $R_{U_s}$ and $R_{Z_s}$ are
the radii at which the quantities are normalized to the solar values.

It is possible that radiation field depends on the azimuthal direction,
because star formation along the spiral arms is enhanced and newly born OB
stars radiate strong UV emission. 
Thus, it seems that there is little UV radiation in the inter-arm regions.
However, Greenawalt et al. (1998) presented deep H$\alpha$ emission-line
images of several spiral galaxies including M51, concluding that a half 
of total H$\alpha$ emission is contributed by the diffuse ionized gas. 
This observation shows that the fluctuation of ionizing radiation field in arm 
and inter-arm regions is not more than twice at their resolution of a few
arcsecs. In the present simulation, the resolution is much lower,
and the arm-interarm UV fields may be assumed to have intensity fluctuation 
less than twice: We here assume that the UV field is uniform along an azimuthal
circle, and depends only on the galactocentric distance.
Although they are a scope beyond the present paper and will be subject for
simulations in the future, variation of radiation fields by 
star-forming arms should be taken into account, and UV radiation transfer 
through the arms, particularly through dark lanes, should be solved, in order
to obtain a more realistic, higher resolution behavior of the 
molecular-fraction across spiral arms.

The interstellar pressure is estimated as $ P \propto \rho (\rm{gas})$, in
the same manner as Honma et al. (1995), and express it by the following
relation:
\begin{equation}
\frac{P}{P_\odot} \sim \frac{\rho}{\rho_\odot} .
\end{equation}
We adopt the following normalizing parameters:
$\rho_{\odot} = 2 {\rm H} \rm{ cm}^{-3}$, $  R_U = R_Z = 5.2$ kpc,
and $ R_{U_s} = R_{Z_s} = 11.5$ kpc.
The latter two parameters are as same as those assumed by Honma et al. (1995).
Although they assumed constant metallicity for NGC 5194, we here assume an
exponential function as above (see e.g. Arimoto et al 1996).

\subsection{Simulated Spiral Molecular Front}

Under these assumptions and initial conditions, we first calculated the time
evolution of the distribution of total gas. 
Fig. 3a shows the density distribution after several galactic rotations. 
The density distribution in the initially uniform-density disk is strongly
disturbed by the oval potential, and high-density arms are 
formed, which evolve into well-developed bisymmetric spiral arms.
In the same time, the global radial density distribution is regulated to 
have a roughly exponentially-decreasing structure.
Inside the bar, straight high-density arms are formed along the axis of 
the bar, as well as many faint spiral-like arms.
We are here interested in the global spiral pattern in order to compare with
the observed gas density distributions, and will not discuss the smaller and
fainter features, which will be finally smeared out when we compare the
result with the observation.

--- Fig, 3a, b, c, d ---

The simulated distribution of gas density is, then, used to calculated
the distribution of gas pressure $P$, which is further combined with 
the assumed distributions of UV radiation field $U$ and the metallicity
$Z$ as given above. Thus, we can calculated the molecular fraction at 
each grid point, and obtain a map of $\fmol$, as well as maps of atomic 
and molecular gases. Fig. 3b, c and d shows snapshots of the thus calculated
maps for HI, \htwo\, and $\fmol$.
In order to compare with the observations, we smooth the map of molecular
fraction by a Gaussian convolution, and compare it with the observation 
of M51 in the same resolution in Fig. 4. 

--- Fig. 4 ---

\section{Discussion}

The simulated results can be summarized as follows.
Detailed arm shapes are not well reproduced, because
the simulation uses a fixed bar-potential.
In order for detailed arm features to be reproduced,
we need self-gravitating $N$-body simulation, including both stars and
gas, is necessary, which is a subject for the future.
However, the simulation in Fig. 3 and 4 shows global agreement with 
the observations, and reproduces the following characteristics of the 
distribution of molecular fraction in M51. 
Around the nucleus $\fmol$ is almost unity, and maintains high values
$r\sim 4$ kpc, where $\fmol> 0.8$.
At $r \sim 4$ to 5 kpc, $\fmol$ decreases drastically to 0.4 or lower,
and becomes nearly zero in the outer region.
It is clear that the transition region, where
$\fmol$ is $\sim$ 0.5 -- 0.6, is distributed in a narrow annulus of radius
of 4.5 kpc, which means that the HI-to-\htwo\ transition occurs rather
abruptly within a narrow range of radius.
This  confirms the radial molecular front (Sofue et al. 1995, Honma et al.
1995).

In addition to the radial MF in the gas disk,
azimuthal fluctuation of molecular fraction is also significant.
At  $r \sim$ 6 kpc, $\fmol$ changes from nearly zero before encountering
the arm to $\sim$ 0.8 within $\sim$ 200 pc in the azimuthal direction in the
spiral arm.
This narrow transition range corresponds to a time scale of $\sim$ 2 Myr.
This indicates that the molecular gas observed along spiral arms is
formed only within $\sim$ 2 Myr.
This time scale is consistent with the H${}_2$ forming time scale
$> 7 \times 10^{5}$ yr predicted by Takahashi \& Williams (2000) using a
molecular dynamics simulation.

On the other hand, the decaying time scale of molecular gas is longer,
according to the more gradual decrease of gas density after the galactic
shock passage.
If we define a molecular arm an armed region with $\fmol>0.5$, the
width of the molecular arm is estimated to be $\sim$ 2.5 kpc in the azimuthal
direction.
This corresponds to a lifetime of molecular gas of about 25 Myr, which is
consistent with the semi-analytically calculated GMCs lifetimes
$\sim$ 40 Myr (Elmegreen 1991).

If we consider the radial variation of metallicity $Z$,
these local (arm) $\fmol$ structures will depend on the galactocentric
distance, even for the same gas density.
Fig. 3 shows that, for the same density of $\sim 5 $ H cm$^{-3}$,
the molecular fraction is $\sim$ 0.5 -- 0.6 in the inner inter-arm region,
while it is 0.3 -- 0.4 in the outer arm regions.
This reflects the fact that the molecular fraction depends not only on the gas
density, but also on the strength of radiation field and metallicity,
particularly the latter.

\vskip 2mm

We are grateful to Dr. M. Honma for helpful advice, and
Mr. J. Koda for discussions on numerical simulations.
The numerical simulations  were performed using the VH-1 code developed by
the numerical astrophysics group at the University of Virginia.

\vskip 10mm
\noindent{\bf References}
\vskip 5mm

\ref Arimoto, N., Sofue, Y., \& Tsujimoto, T. 1996, PASJ 48, 275

\ref Blondin, \& Lufkin, 1993, ApJS, 88, 589

\ref Colella, P., \& Woodward, P. R.,
	1984, Journal of Comp. Phys., 54, 174

\ref Elmegreen, B. G., 1991,
	in ``The Physics of Star Formation and Early Stellar Evolution,
	ed. Lada. C. J., Kylafis, N. D. (Kluwer Academics), 35

\ref Elmegreen, B. 1993, ApJ. 411, 170

\ref Greenawalt, B.,  Walterbos, R. A. M.,  Thilker, D., Hoopes, C. G. 
1998 ApJ, 506, 135.

\ref Imamura, K, \& Sofue, Y., 	1997, AAp, 319, 1

\ref Honma, M., Sofue, Y., \& Arimoto, N., 1995 AAp, 304, 1

\ref  Kuno, N., Nakai, N.,
	Handa, T., \& Sofue, Y., 1994, PASJ, 47, 745

\ref  Kuno, N., \& Nakai, N., 1997, PASJ, 49, 279

\ref  Nakai, N., Kuno, N., Handa, T., \& Sofue, Y., 1994, PASJ, 46, 527

\ref Rots, A. H., Crane, P. C., Bosma, A.,
	Athanassoula, E., \& van der Hulst, J. M., 1990, AJ, 100, 387

\ref Sanders, R. H. 1977 ApJ 217 916

\ref Sofue, Y. 1994 ApJ 423 207

\ref Sofue, Y., Honma, M., \& Arimoto, N., 1995, AAp, 296, 33.

\ref Takahashi, J., \& Williams, D. A., 2000, MNRAS, 314, 27

\ref Toomre, A., 1981, in The Structure and Evolution of Normal Galaxies,
	ed S. M. Fall \& D. Lynden-Bell
	(Cambridge: Cambridge Univ. Press), 111

\newpage

\parindent=0pt

Figure Captions

\vskip 5mm

Fig. 1. (a) HI integrated-intensity map of NGC 5194 (M51) at
$34^{\prime\prime}$ resolution (Rots et al. 1990)(left panel).
The contour levels are 100, 300, 500, \ldots, and 1900 mJy/beam and the
intensity key is shown at the bottom.

(b) $^{12}$CO ($J=1-0$) integrated intensity map of M51 (Nakai et al. 1994),
smoothed to the same spatial resolution ($34^{\prime\prime}$) as that of HI.
The contour interval is 6.6 K km sec${}^{-1}$, and the intensity key is
shown at the bottom.
$10''$ in the maps correspond to linear scale of 466 pc.

\vskip 5mm
Fig. 2. Molecular fraction ($\fmol$) in NGC 5194 (M51).  Contour levels are
$\fmol= 0.1, 0.2, \ldots, 0.9$.

\vskip 5mm
Fig. 3. (a) A snapshot of numerically simulated total density distribution
of interstellar gas.

(b) The same for HI gas.

(c) The same for \htwo\ molecular gas, simulating a CO-line intensity
distribution.

(d) The same for molecular fraction $\fmol$.

The value keys are shown at the bottom.

\vskip 5mm
Fig. 4. Comparison of the simulated molecular fraction (left) smoothed to
the same resolution as that of the observation of M51 (right).

\end{document}